\documentclass[12pt,a4paper]{article}
\usepackage{graphicx}
\usepackage{amsbsy} 
\newcommand{\vs}{\vspace{-0.25cm}}

\begin{document}

\begin{center}
{\Large
\textbf{Chiral four-body interactions in nuclear matter}\footnote{Work supported in 
part by DFG and NSFC (CRC 110).}}

\bigskip
N. Kaiser \\

\bigskip

{\small Physik Department T39, Technische Universit\"{a}t M\"{u}nchen, 
D-85747 Garching, Germany\\

\smallskip

{\it email: nkaiser@ph.tum.de}}

\end{center}

\begin{abstract}
An exploratory study of chiral four-nucleon interactions in nuclear and 
neutron matter is performed. The leading-order terms arising from 
pion-exchange in combination with the chiral $4\pi$-vertex and the chiral 
NN$3\pi$-vertex are found to be very small. Their attractive contribution to the 
energy per particle stays below $0.6\,$MeV in magnitude for densities up to $\rho
=0.4\,$fm$^{-3}$. We consider also the four-nucleon interaction induced by 
pion-exchange and twofold $\Delta$-isobar excitation of nucleons.
For most of the closed four-loop diagrams the occurring integrals over four Fermi 
spheres can  either be solved analytically or reduced to easily manageable 
one- or two-parameter integrals. After summing the individually large contributions 
from 3-ring, 2-ring and 1-ring diagrams of alternating signs, one obtains 
at nuclear matter saturation density $\rho_0=0.16\,$fm$^{-3}$ a moderate contribution 
of $2.35\,$MeV to the energy per particle. The curve $\bar E(\rho)$ rises rapidly with 
density, approximately with the third power of $\rho$. In pure neutron matter the 
analogous chiral four-body interactions lead, at the same density $\rho_n$, to a 
repulsive contribution that is about half as strong. The present calculation 
indicates that long-range multi-nucleon forces, in particular those provided by 
the strongly coupled $\pi N \Delta$-system with its small mass-gap of $293\,$MeV, 
can still play an appreciable role for the equation of state of nuclear and neutron 
matter.  
 \end{abstract}

\medskip


\vspace{-0.3cm}
\section{Introduction and summary}
According to their modern description and construction in chiral effective field 
theory, nuclear forces are organized in hierarchical way \cite{evgeni,hammer}. For 
generic few-body observables the contributions arising from two-nucleon 
interactions are larger than those from three-nucleon forces, and the latter are 
again more important than possible corrections due to four-body forces. By 
constructing  the chiral nucleon-nucleon potential up to next-to-next-to-next-leading 
order (N$^3$LO) one has reached in the effective field theory approach the quality 
of a high-precision NN-potential in reproducing empirical NN-phase shifts and 
deuteron properties \cite{evgeni,hammer,machleidt}. By further lowering the 
resolution scale to $\Lambda \simeq 400\,$MeV the chiral NN-potential can be 
evolved to a low-momentum NN-potential $V_{\rm low-k}$ which becomes nearly 
model-independent and exhibits desirable convergence properties in perturbative 
calculations of many-nucleon systems and infinite nuclear matter 
\cite{vlowkreview,achimnuc,3bodycalc,hebeler}. In chiral effective field theory 
the three-nucleon interaction can be constructed systematically and consistently 
together with the NN-potential \cite{evgeni,hammer}. At leading order it consists 
of a zero-range contact-term, a mid-range $1\pi$-exchange component and a 
long-range $2\pi$-exchange component, where the parameters of the latter occur 
also in the subleading $2\pi$-exchange NN-potential \cite{achimnuc,3bodycalc}. 
The calculation of the subleading chiral three-nucleon force, built up by 
many pion-loop diagrams etc., has been completed in ref.\cite{3bodyn3lo}. 
The first calculation of the neutron matter equation of state with inclusion 
of these subleading chiral three-neutron interactions (at the Hartree-Fock level) 
has been presented recently in ref.\cite{achim}. 

The leading four-nucleon interaction in chiral effective field theory has been 
constructed by Epelbaum in refs.\cite{evgeni4na,evgeni4nb} using the method of 
unitary transformations. The latter method allows to project the dynamics of the 
interacting pion-nucleon system into the purely nucleonic subspace relevant for 
few-nucleon systems below the pion-production threshold. The effect of the 
long-range part of the chiral four-nucleon force on the $^4$He nucleus has been 
estimated in ref.\cite{4bodyhe4}. An additional binding energy of the order 
of a few tenth MeV has been found by computing expectations values with 
totally symmetric wave-functions.  Phenomenological four-nucleon forces arising 
from pion-exchange and twofold $\Delta$-isobar excitation of nucleons have been 
considered in ref.\cite{deltuva} for computations of four-nucleon scattering 
processes ($n$-$^3$H, $p$-$^3$He, $n$-$^3$He, $p$-$^3$H and $d$-$d$). The effect 
of the four-nucleon force on the studied observables was found to be much smaller 
than that of the three-nucleon force, thus confirming the traditional belief in 
the hierarchical order of many-nucleon forces. However, the inclusion of the 
$\Delta$-isobar was not able to resolve some longstanding discrepancies with 
experimental data \cite{deltuva}.

The present paper investigates the effects of chiral four-nucleon interactions 
on the equation of state of homogeneous nuclear and neutron matter. The objective 
is to quantify those particular in-medium processes where four nucleons in the Fermi 
sea interact via the exchange of pions. The special case of pure neutron 
matter is interesting because the presence of neutral pions only leads to reduced 
isospin weight factors. The study of chiral four-body contributions to nuclear 
matter has been initiated in section 5 of ref.\cite{salvatore}, where a particularly 
simple class of 2-ring Hartree diagrams (involving the chiral $4\pi$-vertex and 
the chiral NN$3\pi$-vertex) has been evaluated. With a contribution to the energy 
per particle of less than $0.1\,$MeV for densities up to twice normal nuclear 
matter density, $2\rho_0= 0.32\,$fm$^{-3}$, these long-range four-body 
correlations could be considered as negligibly small. In the present work we 
complete the leading order calculation by evaluating the corresponding  
Fock diagrams with one single closed nucleon ring. In addition to that we 
consider the four-nucleon interaction arising from pion-exchange and twofold 
excitation of nucleons to $\Delta$-isobars. By introducing the direct coupling of 
the pion to the intermediate $\Delta$-isobar ($\Delta\Delta\pi$-vertex) a further 
type of long-range four-nucleon interaction can be generated. Although  motivated
more phenomenologically, these $\Delta$-induced long-range four-nucleon interactions 
occur just as well in the chiral effective field theory framework. It is merely 
a question of how one counts the $\Delta$-nucleon mass splitting ($293\,$MeV), 
whether they are classified as leading order or as next-to-leading order terms. The 
calculations should be trustworthy for Fermi momenta sufficiently below the breakdown 
scale $\sim 500\,$MeV of nuclear chiral effective field theory. 
 
The present paper is organized as follows: In section 2 we evaluate the 
contributions to the energy per particle of isospin-symmetric nuclear matter 
and pure neutron matter from the leading order four-nucleon interaction (involving 
the chiral $4\pi$-vertex and the chiral NN$3\pi$-vertex). Let us note that our 
calculation does not follow the method of unitary transformations of 
ref.\cite{evgeni4nb}. In this scheme iterated two-body and three-body forces give rise 
to induced four-body forces, whereas we restrict ourselves to four-particle irreducible 
diagrams related to ''genuine'' four-nucleon forces (called $V^2_{\rm class-II}$ in 
ref.\cite{evgeni4nb}). A complete study of the chiral four-body contributions to  
neutron matter following the method of unitary transformations has been performed 
recently by the Darmstadt-Ohio group \cite{achim}. Sections 3 and 4 deal
with the $\Delta$-induced four-nucleon interactions mediated by triple pion-exchange.
The different orderings of the pion-couplings in the 
$\Delta$-excitation-deexcitation process are combined into a symmetrized $2\pi$ 
or $3\pi$ contact-vertex proportional to the inverse $\Delta$N mass-splitting 
or its square. Using this convenient short form we present results separately 
for 3-ring, 2-ring and 1-ring diagrams. In all except one case the occurring 
integrals over the product of four Fermi spheres can be either solved 
analytically or reduced to easily manageable one- or two-parameter integrals. 
As a major result we find that the repulsive contribution from the 
dominant 3-ring diagram gets reduced to about 1/3 of its size by the 
(Fock-type) 2-ring and 1-ring diagrams. Altogether, there remains 
at saturation density $\rho_0 =0.16\,$fm$^{-3}$ a contribution of $\bar E(\rho_0)
=2.35\,$MeV to the energy per particle of isospin-symmetric nuclear matter. This 
moderate value lies on a curve for $\bar E(\rho)$ which however rises rapidly with 
density, approximately as $\rho^3$.  In pure neutron matter the analogous chiral 
four-body interactions lead, at the same density $\rho_n$, to a repulsive 
contribution that is about half as strong. 

The present calculation indicates that long-range multi-nucleon forces, in 
particular those provided by the strongly coupled $\pi N \Delta$-system with its 
small mass-gap of $293\,$MeV, are still 
of considerable size. Eventually, when combined with the relatively large attractive 
contributions from the subleading chiral 3N forces found in ref.\cite{achim} for 
pure neutron matter, one can expect cancellations to a certain degree. It remains 
as a future task to demonstrate these expected cancellations at a quantitative 
level  for pure neutron matter and also for isospin-symmetric nuclear matter.   

\section{Leading order four-body terms related to the chiral 
$4\pi$-vertex}
The leading long-range four-nucleon interaction is constructed by connecting the 
four nucleon lines by exchanged pions which couple according to the chiral pion-pion 
interaction. Since the off-shell chiral $4\pi$-vertex is involved in the process 
this contribution alone depends on the choice of the interpolating pion-field  
\cite{evgeni4nb} and thus is not unique. It has to be supplemented by the four-nucleon 
interaction generated additionally by the chiral NN$3\pi$-vertex, where the three 
pions emitted from one nucleon are absorbed on each of the other three nucleons. 
The sum of both pieces is actually representation-independent and thus gives 
rise to unique and physically meaningful results. The isovector (Weinberg-Tomozawa) 
NN$\pi\pi$-vertex is proportional to the energies of the exchanged pions, which in 
the present application are differences of nucleon kinetic energies. According to 
this property the four-nucleon interaction generated by two (Weinberg-Tomozawa) 
NN$\pi\pi$-vertices is a relativistic $1/M^2$-correction. This observed suppression 
is equivalent to the statement $V_{\rm class-III}=0$ in ref.\cite{evgeni4nb}. Note that 
we do not follow here the method of unitary transformations \cite{evgeni4nb}, but 
consider only ''irreducible'' diagrams, which would be commonly classified  as 
''genuine'' four-nucleon interactions. These pieces are called $V^2_{\rm class-II}$ 
in ref.\cite{evgeni4nb}.   

\begin{figure}
\begin{center}
\includegraphics[scale=0.6,clip]{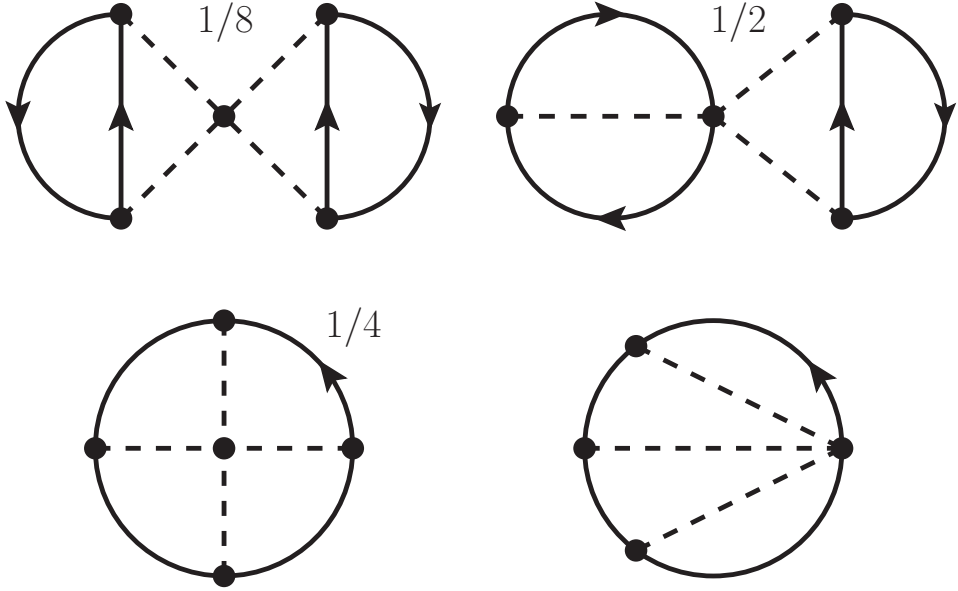}
\end{center}
\vspace{-.8cm}
\caption{2-ring and 1-ring diagrams related to the leading order (genuine) chiral 
four-nucleon interaction. The combinatorial factor of each diagram is specified, 
if it is unequal to $1$.}
\end{figure}

Fig.\,1 shows the diagrams one obtains by closing the four nucleon lines to 
either two rings or one ring. The combinatoric factors of these diagrams are $1/8$, 
$1/2$, $1/4$ and $1$, respectively. Diagram with more rings are trivially zero, 
due to a  vanishing spin-trace. For the 2-ring (Hartree) diagrams the integrals 
over the four Fermi spheres factorize and can be readily solved by using the 
master formula eq.(29) in the appendix. One finds the following contribution to the 
energy per particle of isospin-symmetric nuclear matter:
\begin{equation}\bar E(\rho)=  {9g_A^4 m_\pi^7 u\over (4\pi f_\pi)^6}
\bigg[u^2-{1\over 2}-2u \arctan 2u+\bigg(1+{1\over 8u^2}\bigg)\ln(1+4u^2)
\bigg]^2\,, \end{equation}
with the dimensionless variable $u = k_f/m_\pi$. The nucleon density $\rho$ is 
related to the Fermi momentum $k_f$ in the usual way, $\rho = 2k_f^3/3\pi^2$. The 
occurring parameters are: $g_A=1.3$ (nucleon axial-vector coupling constant),  
$f_\pi=92.4\,$MeV (pion decay constant) and  $m_\pi=135\,$MeV (neutral pion mass). 
Note that after adding both 2-ring diagrams in Fig.\,1 the only remainder of the 
chiral $\pi\pi$-interaction is a constant factor $-3m_\pi^2/f_\pi^2$. This feature 
has ultimately lead to the expression with a complete square in eq.(1). The 
analogous chiral four-body contribution in pure neutron matter is also of interest. 
In this case only neutral pions are present, which leads to a reduced isospin 
factor. Doing the calculation one finds from the sum of both 2-ring diagrams in 
Fig.\,1 the following contribution to the energy per particle of pure neutron matter:  
\begin{equation}\bar E_n(\rho_n)=  -{3g_A^4 m_\pi^7 u\over 2(4\pi f_\pi)^6}
\bigg[u^2-{1\over 2}-2u \arctan 2u+\bigg(1+{1\over 8u^2}\bigg)\ln(1+4u^2)
\bigg]^2\,, \end{equation}
with $u = k_n/m_\pi$. The neutron density $\rho_n$ is related to the neutron Fermi 
momentum $k_n$ by $\rho_n = k_n^3/3\pi^2$. In order to simplify the notation, we 
make from here on the agreement that in all formulas for $\bar E_n(\rho_n)$ the 
dimensionless variable $u$ has the meaning $u = k_n/m_\pi$, while in all
formulas for $\bar E(\rho)$ it keeps its original meaning $u=k_f/m_\pi$. 
\begin{figure}
\begin{center}
\includegraphics[scale=0.5,clip]{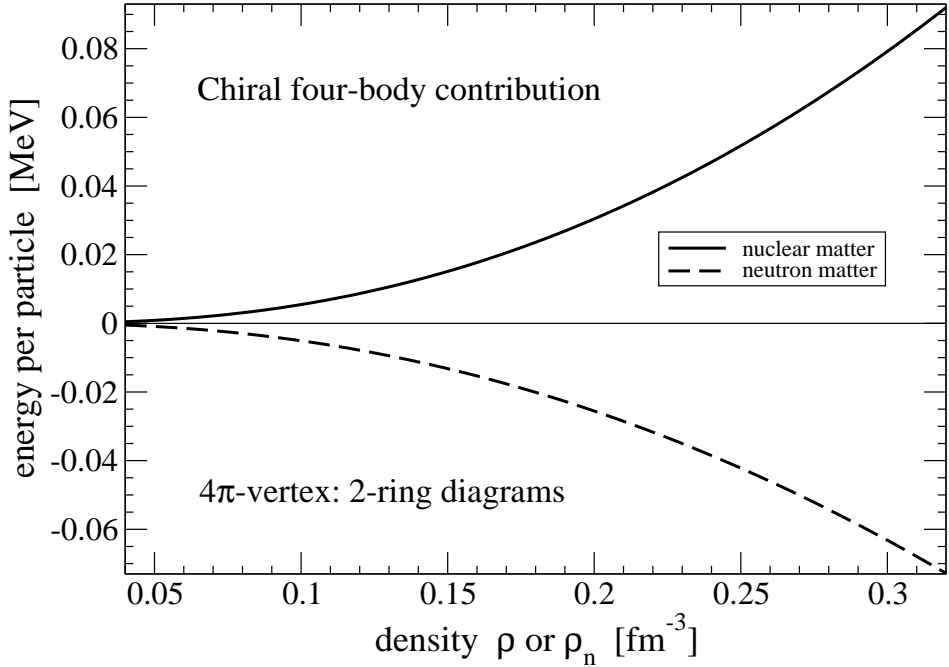}
\end{center}
\vspace{-.8cm}
\caption{Chiral four-body contributions to the energy per particle of nuclear and 
neutron matter arising from leading order 2-ring diagrams.}
\end{figure}

The full and dashed line in Fig.\,2 show these chiral four-body contributions to the 
equation of state of nuclear and neutron matter in the density region $0.04\,$fm$^
{-3}< \rho, \rho_n <  0.32\,$fm$^{-3}$. One observes very small repulsive or 
attractive contributions which do not exceed $0.1\,$MeV in magnitude. Note that  
at equal densities, $\rho_n=\rho$, the isospin reduction factor $-1/6$ for pure 
neutron matter gets effectively compensated by the increase of the Fermi momentum 
$k_n$ (by a factor $2^{1/3}$).
   
Next, we turn to the evaluation of the 1-ring (Fock) diagrams shown in the lower 
part of Fig.\,1. In order to keep the symmetry of the left diagram (with a 
$4\pi$-vertex), it is advantageous to add four versions of the right diagram 
(with a NN$3\pi$-vertex), each then weighted by a factor $1/4$. Putting together 
all pieces emerging from spin- and isospin-traces, one finds the following 
contribution to the energy per particle of isospin-symmetric nuclear matter:    
\begin{eqnarray}\bar E(\rho)&=& {3g_A^4 \over 16f_\pi^6 \rho} \int\limits_{
|\vec p_j|<k_f}\!\!\!{d^{12}p \over (2\pi)^{12}}\, {1\over m_\pi^2+\vec q_1^{\,2}}\,
{1\over m_\pi^2+\vec q_2^{\,2}}\,{1\over m_\pi^2+\vec q_3^{\,2}}\, {1\over 
m_\pi^2+(\vec q_1+\vec q_2+\vec q_3)^{2}}\nonumber \\ && \times \bigg\{ -8[
\vec q_1\!\cdot\!(\vec q_2\!\times\! \vec q_3)]^2+m_\pi^2\Big[-4\vec q_1^{\,2}\,
\vec q_3^{\,2}-4\vec q_1^{\,2}(\vec q_2^{\,2}+\vec q_1\!\cdot\! \vec q_3)\nonumber 
\\ && +2\vec q_1^{\,2}\,\vec q_2\!\cdot\!(2\vec q_1-5\vec q_3)+8(\vec q_1\!
\cdot\!\vec q_2)^2 +\vec q_1\!\cdot\!\vec q_3
\, \vec q_2\!\cdot\!(8\vec q_1+3\vec q_2)-6\vec q_1\!\cdot\!\vec q_2\,
\vec q_2\!\cdot\!\vec q_3\Big]\bigg\}\,, \end{eqnarray}
with momentum transfers $\vec q_1 = \vec p_1 -\vec p_4$,   $\vec q_2 = \vec p_2 
-\vec p_1$ and  $\vec q_3 = \vec p_3 -\vec p_2$. In the case of pure neutron matter, 
where the chiral four-body interaction is mediated by neutral pions only, one gets 
from the sum of the 1-ring diagrams:
\begin{eqnarray}\bar E_n(\rho_n)&=& {g_A^4 m_\pi^2\over 32f_\pi^6 \rho_n} \int
\limits_{|\vec p_j|<k_n}\!\!\!{d^{12}p \over (2\pi)^{12}}\, {1\over m_\pi^2+\vec 
q_1^{\,2}}\,{1\over m_\pi^2+\vec q_2^{\,2}}\,{1\over m_\pi^2+\vec q_3^{\,2}}\, 
\nonumber \\ && \times {1\over m_\pi^2+(\vec q_1+\vec q_2+\vec q_3)^{2}}
\Big\{\vec q_2^{\,2}\, \vec q_1\!\cdot\!\vec q_3-2 \vec q_2\!\cdot\!
\vec q_3 (\vec q_1+\vec q_2)\!\cdot\!\vec q_1\Big\}\,. \end{eqnarray}
Note that we have exploited the symmetry under the interchange $\vec q_1 
\leftrightarrow \vec q_3$ in order to simplify the expressions in the curly 
brackets of eqs.(3,4). The occurrence of a product of four different 
pion-propagators prohibits any substantial analytical reduction of these 
12-dimensional integrals over the product of four Fermi spheres. Note also that 
there is a marked difference between $\bar E(\rho)$ in eq.(3) and 
$\bar E_n(\rho_n)$ in eq.(4). After adding both 1-ring diagrams the result for 
pure neutron matter becomes proportional to $m_\pi^2$, while there remains a 
finite term $-8[\vec q_1\!\cdot\!(\vec q_2\!\times\! \vec q_3)]^2$ for 
isospin-symmetric nuclear matter even in the chiral limit $m_\pi=0$. For comparison, 
the sum of the 2-ring (Hartree) diagrams has generated the $m_\pi^2$ factor in 
both cases. We note as an aside that in the chiral limit, $m_\pi=0$, eq.(3) leads to 
the result $\bar E(\rho)\simeq -1.13\,g_A^4 k_f^7/(4\pi f_\pi)^6$ with a simple seventh 
power dependence on the Fermi momentum $k_f$.     

\begin{figure}
\begin{center}
\includegraphics[scale=0.5,clip]{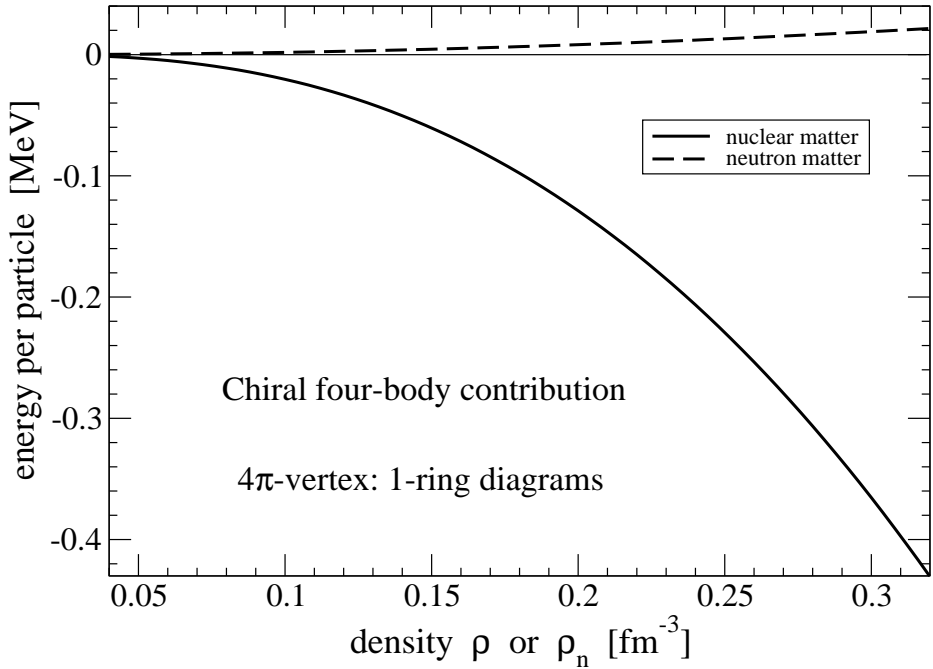}
\end{center}
\vspace{-.8cm}
\caption{Chiral four-body contributions to the energy per particle of nuclear and 
neutron matter arising from leading order 1-ring diagrams.}
\end{figure}

The full and dashed line in Fig.\,3 show the numerical results for the 1-ring (Fock) 
contributions as a function of the density, $\rho$ or $\rho_n$. In the case of pure
neutron matter one finds the usual pattern with the Fock contribution having 
opposite sign and reduced spin-weight in comparison to the Hartree contribution 
(shown in Fig.\,2). In contrast to this the Fock contribution for isospin-symmetric 
nuclear matter is about a factor of five larger (and of opposite sign) than the 
Hartree contribution. This reversed ordering comes from the fact that the Hartree 
contribution eq.(1) is ``accidentally'' suppressed by a factor $(m_\pi/k_f)^2$, 
while the Fock contribution remains finite in the chiral limit $m_\pi=0$. 
The Hartree terms shown in Fig.\,2 are therefore not representative for the size 
of chiral four-body contributions. Irrespective of that exceptional behavior, 
the attractive chiral four-body contributions represented by the diagrams in 
Fig.\,1 stay in total below $0.6$\,MeV for densities up to $\rho=0.4\,$fm$^{-3}$. 
For this reason they can be considered as negligibly small. 

Let us remark that in the case of pure neutron matter the method of unitary 
transformations \cite{evgeni4nb} provides just one additional leading order 
chiral four-body interaction (called $V^{a}$ in ref.\cite{achim}). It has been 
computed in ref.\cite{achim} with the result that its contribution to the energy 
per particle $\bar E_n(\rho_n)$ reaches about $-0.25\,$MeV at a density of 
$\rho_n=0.2\,$fm$^{-3}$.  

\section{Single delta-isobar excitation on two nucleons}
In this section we consider a more relevant type of long-range four-nucleon 
interaction \cite{deltuva}. It arises from triple pion-exchange and the 
excitation of two nucleons into virtual $\Delta$-isobars. The corresponding 
four-nucleon interaction is depicted by the left diagram in Fig.\,4. For an 
efficient treatment it is advantageous to combine the direct and crossed 
$\pi N \to \Delta\to \pi N$ transition into a contact-vertex for the absorbed 
and emitted pion. In the present application the energy dependence of the 
$\Delta$-propagator $i(\omega -\Delta)^{-1}$ can be neglected since the energies 
$\omega$ carried by the virtual pions are differences of small nucleon kinetic 
energies. With this valid approximation the pertinent contact-vertex reads:
\begin{equation}{i g_A^2 \over f_\pi^2 \Delta} \bigg[ \delta_{ab}\, 
\vec q_a\cdot\vec q_b -{1\over 4} \epsilon_{abc} \tau_c \, \vec \sigma
\cdot(\vec q_a\times\vec q_b) \bigg]\,, \end{equation}
where $\vec q_a$ is an ingoing and $\vec q_b$ is an outgoing pion-momentum. 
We have inserted in eq.(5) already the empirically well-satisfied coupling 
constant ratio $g_{\pi N\Delta}/g_{\pi NN}=3/\sqrt{2}$, and $\Delta = 293\,$MeV denotes 
the mass-splitting between the $\Delta$-isobar and the nucleon. Note that the 
scale $\Delta = 293\,$MeV is comparable to the Fermi momentum $k_{f0}=263\,$MeV at 
nuclear matter saturation density $\rho_0=0.16\,$fm$^{-3}$. At $\rho_n =2\rho_0$ the 
neutron Fermi momentum $k_n = 418\,$MeV exceeds the scale $\Delta = 293\,$MeV
already considerably. For the in-medium interaction the momenta of the exchanged 
pions vary over the range $0<|\vec q_{a,b}| < 2k_{f,n}$, but the included 
high-momentum part receives very little weight in the Fermi sphere integrals 
(as demonstrated in the appendix). When working with a NN$\pi\pi$ contact-vertex 
the task of enumerating all topologically distinct in-medium diagrams gets simplified 
a lot. We turn now to the analytical evaluation of the pertinent four-loop in-medium 
diagrams representing the energy density, $\rho \bar E(\rho)$ or 
$\rho_n \bar E_n(\rho_n)$. 

\begin{figure}
\begin{center}
\includegraphics[scale=0.8,clip]{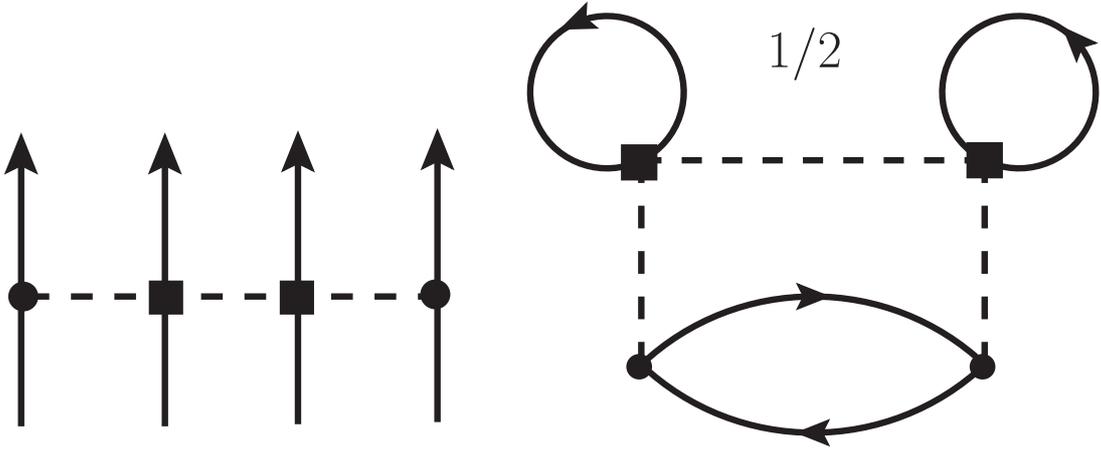}
\end{center}
\vspace{-.8cm}
\caption{Left: Long-range four-nucleon interaction with twofold $\Delta$-isobar 
excitation. Right: Corresponding 3-ring Hartree diagram with combinatorial factor 
$1/2$.}
\end{figure}

\subsection{Three-ring diagram}
This simplest closed in-medium diagram resulting from the $\Delta$-induced 
four-nucleon interaction is the 3-ring diagram shown in the right part of Fig.\,4. 
The upper two loops produce each a factor of density, $\rho$ or $\rho_n$, and the 
remaining integral over two Fermi spheres is readily solved by using eq.(29). One 
finds the following contributions to the energy per particle of isospin-symmetric 
nuclear matter:  
\begin{equation}\bar E(\rho)= {g_A^6 m_\pi^9 u^3\over (2\pi f_\pi)^6\Delta^2}
\bigg\{{16 u^6\over 9}-12u^4 +{20u^2\over 3}+{70 u^3 \over 3} 
 \arctan 2u-\bigg(12u^2+{5\over 3}\bigg)\ln(1+4u^2)\bigg\}\,, \end{equation}
and pure neutron matter:
\begin{equation}\bar E_n(\rho_n)= {g_A^6 m_\pi^9 u^3\over (2\pi f_\pi)^6\Delta^2}
\bigg\{{4 u^6\over 27}-u^4 +{5u^2\over 9}+{35 u^3 \over 18} 
 \arctan 2u-\bigg(u^2+{5\over 36}\bigg)\ln(1+4u^2)\bigg\}\,, \end{equation}
where the latter differs from the former only by an isospin reduction factor $1/12$ 
(at equal Fermi momenta, $k_n=k_f$). 

Fig.\,5 shows the four-body contributions to the energy per particle of 
nuclear and neutron matter as they arise from the 3-ring diagram with twofold
$\Delta$-excitation. For isospin-symmetric nuclear matter (see full line) one 
obtains a sizeable repulsion which reaches a value of $74\,$MeV at twice normal 
nuclear matter density $2\rho_0=0.32\,$fm$^{-3}$. At saturation density 
$\rho_0=0.16\,$fm$^{-3}$ the corresponding value $\bar E(\rho_0)=7.42\,$MeV is ten 
times smaller, but this amounts still to almost half of the empirical nuclear matter 
binding energy $\bar E_0 \simeq -16\,$MeV. The dashed curve in Fig.\,5 for pure 
neutron matter lies about $20\%$ below the full curve for isospin-symmetric nuclear 
matter. This feature demonstrates that the isospin reduction factor $1/12$ in 
eq.(7) gets to a large extent compensated by the increase of the neutron Fermi 
momentum $k_n$.

With the analytical formulas eqs.(6,7) and the corresponding numerical results 
displayed in Fig.\,5 the size of long-range $\Delta$-induced four-nucleon 
contributions in nuclear and neutron matter is set. From here on the interesting 
question is, how much of a reduction will be provided by the Fock-type diagrams with 
fewer rings.  

\begin{figure}
\begin{center}
\includegraphics[scale=0.5,clip]{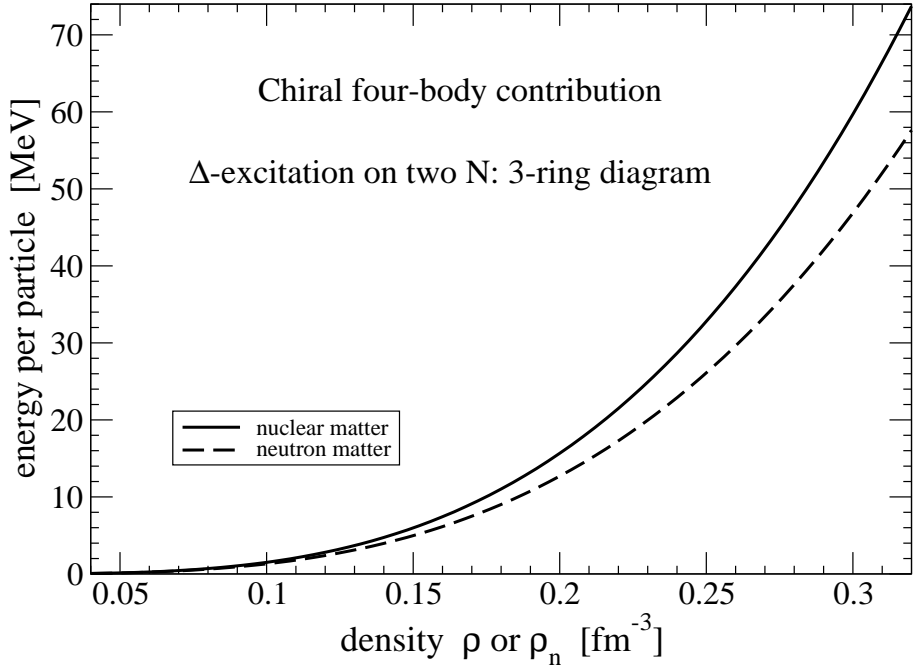}
\end{center}
\vspace{-.8cm}
\caption{$\Delta$-induced chiral four-body contributions to the energy per 
particle of nuclear and neutron matter: 3-ring diagram.}
\end{figure}

\subsection{Two-ring diagrams}
The 2-ring diagrams resulting from the chiral four-nucleon interaction with 
twofold $\Delta$-isobar excitation are shown in Fig.\,6. The last two diagrams 
(marked by $0$) vanish identically. The upper diagram of this subset involves a 
central pion which carries zero-momentum and therefore  does not couple according 
to the contact-vertex eq.(5). In order not to vanish in the first step, the 
spin-traces in the lower diagram select from the contact-vertex eq.(5) the 
spin-dependent part. Furthermore, by taking into account momentum conservation at each 
vertex one obtains a scalar triple product of three linearly dependent vectors 
($\vec q_1+\vec q_2+\vec q_3 =  \vec 0$), which vanishes identically. 

\begin{figure}
\begin{center}
\includegraphics[scale=0.7,clip]{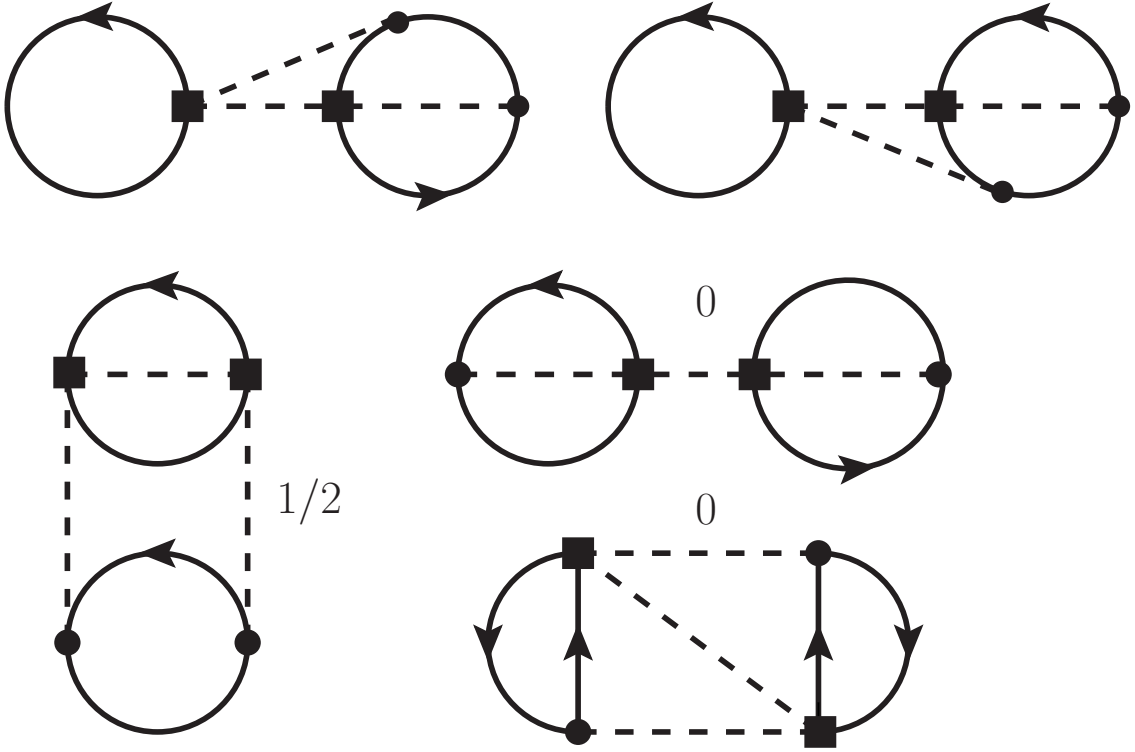}
\end{center}
\vspace{-.8cm}
\caption{2-ring diagrams resulting from the chiral four-nucleon interaction with 
twofold $\Delta$-isobar excitation.}
\end{figure}

Obviously, the upper two (topologically distinct) diagrams in Fig.\,6 give rise to 
equal contributions. The left loop produces just a factor of density and the Fermi 
sphere integrals over the pion-propagators and momentum-dependent interactions on 
the right can be factorized with the help of tensors. Performing this strategy, as 
outlined in eq.(30) of the appendix, one ends up with the following expressions 
for the contributions to the energy particle of nuclear and neutron matter:     
\begin{equation}\bar E(\rho)= -{g_A^6 m_\pi^9 \over 4(2\pi f_\pi)^6 \Delta^2}
\int_0^u\!dx \Big[2 G_S(x) H_S(x)+G_T(x)H_T(x)\Big]\,,\end{equation}
\begin{equation}\bar E_n(\rho_n)= -{g_A^6 m_\pi^9 \over 24(2\pi f_\pi)^6\Delta^2}
\int_0^u\!dx \Big[G_S(x) H_S(x)+2G_T(x)H_T(x)\Big]\,.\end{equation}
Here, we have introduced the auxiliary functions:
\begin{eqnarray} G_S(x) &=&{4ux \over 3}( 2u^2-3)+ 4x\Big[\arctan(u+x) 
+\arctan(u-x)\Big]\nonumber \\ && + (x^2-u^2-1) \ln{1+(u+x)^2\over  1+(u-x)^2} 
\,,\end{eqnarray}
\begin{eqnarray} G_T(x) &=& {ux\over 6}(8u^2+3x^2)-{u\over
2x} (1+u^2)^2  + {1\over 8} \bigg[ {(1+u^2)^3 \over x^2} \nonumber \\ && 
-x^4+(1-3u^2)(1+u^2-x^2)\bigg] \ln{1+(u+x)^2\over  1+(u-x)^2} \,,\end{eqnarray}
defined by the tensorial Fermi sphere integral in eq.(30) of the appendix. The 
other auxiliary functions $H_{S,T}(x)$ occurring in eqs.(8,9) are closely related 
to $G_{S,T}(x)$:
\begin{equation}H_S(x)={5\over2} G_S(x)+2u x(1-2u^2)+{1\over 2}(u^2-x^2-1) 
\ln{1+(u+x)^2\over  1+(u-x)^2}\,,\end{equation}
\begin{equation}H_T(x)=G_T(x)+ {ux\over 2}-{3u\over2x} (1+u^2)  + {1\over 8} 
\bigg[ {3\over x^2}(1+u^2)^2 +2-2u^2-x^2\bigg] \ln{1+(u+x)^2\over  1+(u-x)^2} 
\,.\end{equation}
The lower left 2-ring diagram in Fig.\,6 (with a combinatorial factor $1/2$)
is more difficult to evaluate. In order to facilitate an analytical treatment    
one chooses the momentum transfers carried by the exchanged pions as the basic 
variables and works with integrals over shifted Fermi spheres. In the end one 
arrives at a double-integral representation for the contributions to the energy 
particle of nuclear and neutron matter:   
\begin{eqnarray}\hspace{-0.7cm}\bar E(\rho)&=& {g_A^6 m_\pi^9 \over (2\pi f_\pi)^6
\Delta^2 u^3}\!\int_0^u\! dx\!\int_0^u\!dy \,{x^3(u^2-y^2) \over (1+4x^2)^2} 
(u-x)^2(2u+x) \Bigg\{7xy (1+4y^2)^2 \nonumber \\ && + 16 x^3\bigg[ 10y
-{136y^3 \over 3}-7x^2y-5 \arctan(2x+2y)+5 \arctan(2x-2y) \bigg]
 \nonumber \\ && +{1\over 16}(4x^2-4y^2-1)\Big[7(1+4y^2)^2 +8x^2(14x^2-28y^2-13)
\Big] \ln{1+4(x+y)^2\over 1+4(x-y)^2} \Bigg\}\,, \end{eqnarray}
\begin{eqnarray}\bar E_n(\rho_n)&=& {g_A^6 m_\pi^9 \over 3(2\pi f_\pi)^6
\Delta^2 u^3}\!\int_0^u\! dx\!\int_0^u\!dy\,{x^3 (u^2-y^2)\over 
(1+4x^2)^2} (u-x)^2(2u+x)\nonumber \\ && \times \Bigg\{ 4xy(1+4y^2)^2 + 32x^3
\bigg[ 2y-{32y^3 \over 3}-2x^2y- \arctan(2x+2y) \nonumber \\ && +\arctan(2x-2y) 
\bigg]+{1\over 4}(4x^2-4y^2-1)^3 \ln{1+4(x+y)^2\over 1+4(x-y)^2} \Bigg\}\,. 
\end{eqnarray}
By inspecting the first lines in eqs.(14,15) one realizes that the master integral 
eq.(29) has been employed. In addition to that, the following reduction formula has 
helped to eliminate a complicated angular integral:   
\begin{equation} \int_0^u\!dp\!\int_{-1}^1\,dz\, p^2 F\Big(p z+\sqrt{u^2-p^2
(1-z^2)}\Big)= \int_0^u\!dy\,(u^2-y^2)F(2y)\,,\end{equation}
where the argument of $F(\dots)$ on the left hand side gives the distance to the 
boundary of a shifted Fermi sphere, with $p$ being the displacement of the center. 
We have checked numerically the analytical expressions in eqs.(14,15) against 
the original (untreated) four Fermi sphere integrals and found good agreement. 
The advantage of the double-integral representations for $\bar E(\rho)$ and 
$\bar E_n(\rho_n)$ in eqs.(14,15) is that these can be evaluated with much higher 
numerical precision.  

The full and dashed line in Fig.\,7 show the $\Delta$-induced chiral four-body 
contributions to the energy per particle of nuclear and neutron matter as they 
arise from the  2-ring diagrams. The first part given by eqs.(8,9) is typically 
a factor 3 to 4 larger than the second part given by eqs.(14,15) and both 
pieces come with the same sign.  By comparison with Fig.\,5 one observes that 
the individually large contributions from the 3-ring and 2-ring diagrams nearly 
cancel each other, leaving a small repulsive (attractive) remainder for nuclear
(neutron) matter. In magnitude this remainder is only about $1/7$ of the starting 
values. As a consequence of this feature the final result for the 
$\Delta$-induced chiral four-body contributions will be determined in a crucial
way by the 1-ring diagrams. 

\begin{figure}
\begin{center}
\includegraphics[scale=0.5,clip]{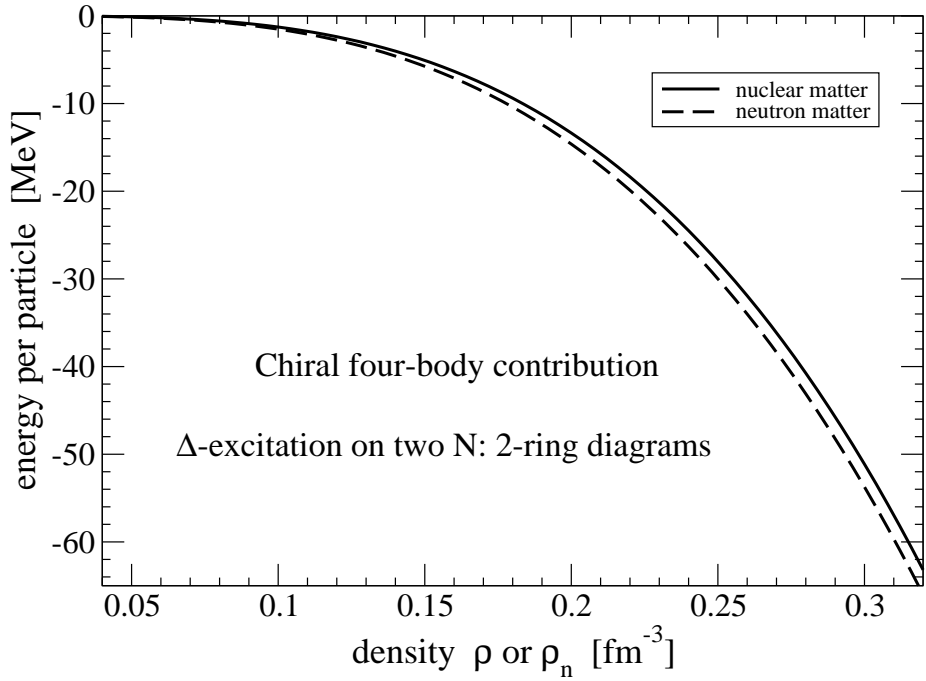}
\end{center}
\vspace{-.8cm}
\caption{$\Delta$-induced chiral four-body contributions to the energy per 
particle of nuclear and neutron matter: 2-ring diagrams.}
\end{figure}

\subsection{One-ring diagrams}

\begin{figure}
\begin{center}
\includegraphics[scale=0.55,clip]{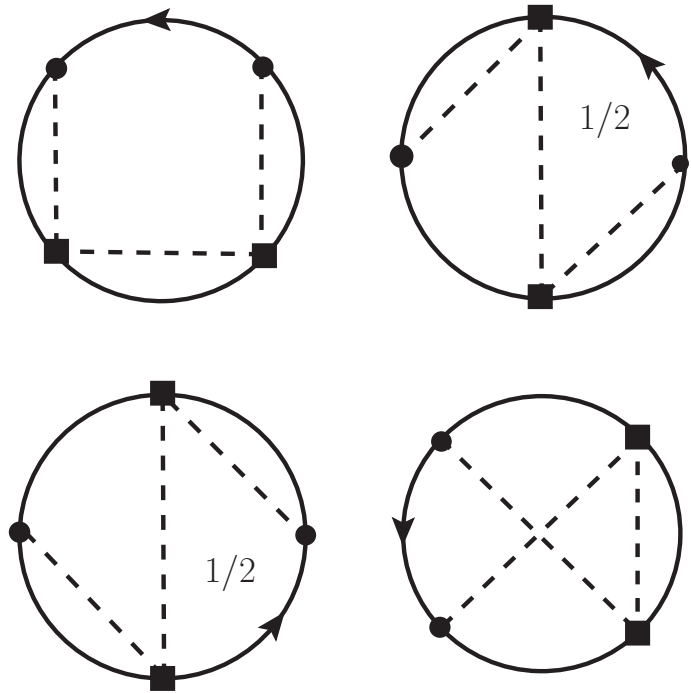}
\end{center}
\vspace{-.8cm}
\caption{1-ring diagrams resulting from the chiral four-nucleon interaction with 
twofold $\Delta$-isobar excitation.}
\end{figure}

The 1-ring diagrams resulting from the chiral four-nucleon interaction with 
twofold $\Delta$-isobar excitation are shown in Fig.\,8. For the first diagram 
with a U-shaped pion-line three of the four occurring Fermi sphere integrals over the
pion-propagators are structurally equivalent and with inclusion of the 
momentum-dependent interactions they factorize with the help of tensors. In the end
only one single radial integral remains and the corresponding contributions 
to the energy per particle of nuclear and neutron matter read:       
\begin{equation}\bar E(\rho)= {g_A^6 m_\pi^9 \over 8(4\pi f_\pi)^6
\Delta^2 u^3}\int_0^u\!dx \,{1\over x}\Big[8 G^3_S(x)+ 9G_S(x) G_T^2(x)+ 
G_T^3(x)\Big]\,, \end{equation}
\begin{equation}\bar E_n(\rho_n)= {g_A^6 m_\pi^9 \over 12(4\pi f_\pi)^6\Delta^2 
u^3}\int_0^u\!dx \,{1\over x}\Big[G^3_S(x)+ 6G_S(x) G_T^2(x)+ 2G_T^3(x)\Big]\,, 
\end{equation}
with the auxiliary functions $G_{S,T}(x)$ given in eqs.(10,11). The next two 
(topologically distinct) 1-ring diagrams in Fig.\,8 have a pion-line running zig-zag. 
Since both give equal contributions their combinatorial factor $1/2$ gets 
effectively compensated. Again, the Fermi sphere integrals over the 
pion-propagators associated to the short pion-lines factorize via tensors. For 
the remaining integral over two Fermi spheres $|\vec p_{1,2}|<k_f$ the angular 
integrations can be carried out and in this procedure the third pion-propagator 
introduces the logarithmic function:       
\begin{equation}L = \ln{1+(x+y)^2 \over 1+(x-y)^2} \,. \end{equation} 
Putting all pieces together and exploiting the symmetry under the interchange 
$x \leftrightarrow y$, one finds the following double-integral representations 
for the contributions of the zig-zag 1-ring diagrams to the energy per particle 
of nuclear and neutron matter:
\begin{eqnarray}\bar E(\rho)&=& {g_A^6 m_\pi^9 \over (4\pi f_\pi)^6\Delta^2 u^3}
\int_0^u\!dx\!\int_0^u\!dy \,\Bigg\{ 2 G_S(x) G_S(y)\Big[4x y-L\Big] + G_S(x) 
G_T(y) \nonumber \\ && \times \bigg[ 5x y-{3x \over y}(1+x^2) +{L\over 4}
\bigg({3\over y^2}(1+x^2)^2 -2-6x^2 +3y^2\bigg) \bigg] +
G_T(x) G_T(y)  \nonumber \\ && \times \bigg[ {x y \over 8} +{3 \over 8y}\bigg( 
{3 \over 2x} +2x-x^3 \bigg) +{L\over 32} \bigg({3 \over y^2}\Big( x^4-x^2-5 
-{3\over 2 x^2}\Big) -1-3x^2 \bigg)\bigg] \Bigg\}\,, \end{eqnarray}
\begin{eqnarray}\bar E_n(\rho_n)&=& {g_A^6 m_\pi^9 \over 6(4\pi f_\pi)^6\Delta^2 
u^3}\int_0^u\!dx\!\int_0^u\!dy \,\Bigg\{G_S(x) G_S(y)\Big[4x y-L\Big] + G_S(x) 
G_T(y) \nonumber \\ && \times \bigg[ 10x y-{6x \over y}(1+x^2) +{L\over 2}
\bigg({3\over y^2}(1+x^2)^2 -2-6x^2 +3y^2\bigg) \bigg] +
G_T(x) G_T(y)  \nonumber \\ && \times \bigg[ x y  +{3 \over y}\bigg( 
{3 \over 2x} +2x-x^3 \bigg) +{L\over 4} \bigg({3 \over y^2}\Big( x^4-x^2-5 
-{3\over 2 x^2}\Big) -1-3x^2 \bigg)\bigg] \Bigg\}\,. \end{eqnarray}
The last 1-ring diagram in Fig.\,8 involves crossed pion lines. The assignment of 
momenta to the virtual pions is now such that the previous factorization procedure 
does not work anymore.  From the four Fermi sphere integrals only the one associated 
to the nucleon line between the contact-vertices can be performed analytically and the 
contributions to the energy per particle of nuclear and neutron matter read:
\begin{eqnarray}\bar E(\rho)&=& {3g_A^6 \over 32f_\pi^6 \Delta^2 u^3} 
\int\limits_{|\vec p_j|<k_f}\!\!\!{d^9p \over (2\pi)^9}\, {\vec q_1^{\,2}\,
\over (m_\pi^2+\vec q_1^{\,2})(m_\pi^2+\vec q_2^{\,2}) |\vec \eta\,|} \nonumber 
\\ && \times\bigg\{\vec q_2^{\,2}\Big[G_S(|\vec \eta\,|)-G_T(|\vec \eta\,|)
\Big] + {3\over \vec \eta^{\,2}}(\vec \eta\cdot\vec q_2)^2\, G_T(|\vec \eta\,|)
\bigg\} \,,  \end{eqnarray}
\vspace{-0.3cm}
\begin{eqnarray}\bar E_n(\rho_n)&=& {g_A^6 \over 32f_\pi^6 \Delta^2 u^3} 
\int\limits_{|\vec p_j|<k_n}\!\!\!{d^9p \over (2\pi)^9}\, {\vec q_1\cdot 
\vec q_2\over (m_\pi^2+\vec q_1^{\,2})(m_\pi^2+\vec q_2^{\,2}) |\vec \eta\,|} 
\nonumber \\ && \times\bigg\{\vec q_1\cdot \vec q_2\Big[G_S(|\vec \eta\,|)
-G_T(|\vec \eta\,|)\Big] + {3\over \vec \eta^{\,2}}\,\vec \eta\cdot\vec q_1\,
\vec \eta\cdot\vec q_2 \,G_T(|\vec \eta\,|)\bigg\} \,,  \end{eqnarray}
with momentum transfers $\vec q_1=\vec p_1-\vec p_3$,  $\vec q_2=\vec p_2-\vec p_3$ 
and the vector $\vec \eta=(\vec p_1+\vec p_2-\vec p_3)/m_\pi$.

\begin{figure}
\begin{center}
\includegraphics[scale=0.486,clip]{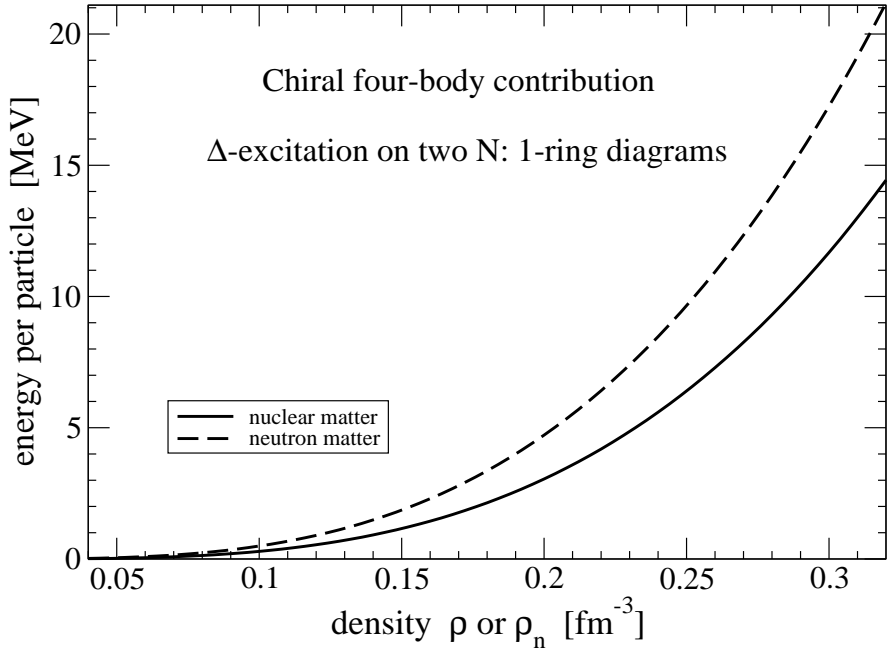}
\end{center}
\vspace{-.8cm}
\caption{$\Delta$-induced chiral four-body contributions to the energy per 
particle of nuclear and neutron matter: 1-ring diagrams.}
\end{figure}

The full and dashed line in Fig.\,9 show the $\Delta$-induced chiral four-body 
contributions to the energy per particle of nuclear and neutron matter as they arise
from the 1-ring diagrams. The three pieces (with U-shaped, zig-zag and crossed 
pion-lines) provide approximately equal amounts. As expected the sign of 
$\bar E(\rho)$ and $\bar E_n(\rho_n)$ is opposite to the contributions from 
the 2-ring diagrams shown in Fig.\,7. Although the magnitude of the repulsive 
1-ring contributions is substantially reduced (by a factor 3 to 4) in comparison to 
the previous terms, these are by no means small. Such strongly with density rising 
four-body correlations, which reach values of $14.4\,$MeV or $21.2\,$MeV at 
$2\rho_0=0.32\,$fm$^{-3}$, affect the equation of state of nuclear or neutron matter 
appreciably.   
 
\section{Double delta-isobar excitation on one nucleons}
\begin{figure}
\begin{center}
\includegraphics[scale=0.515,clip]{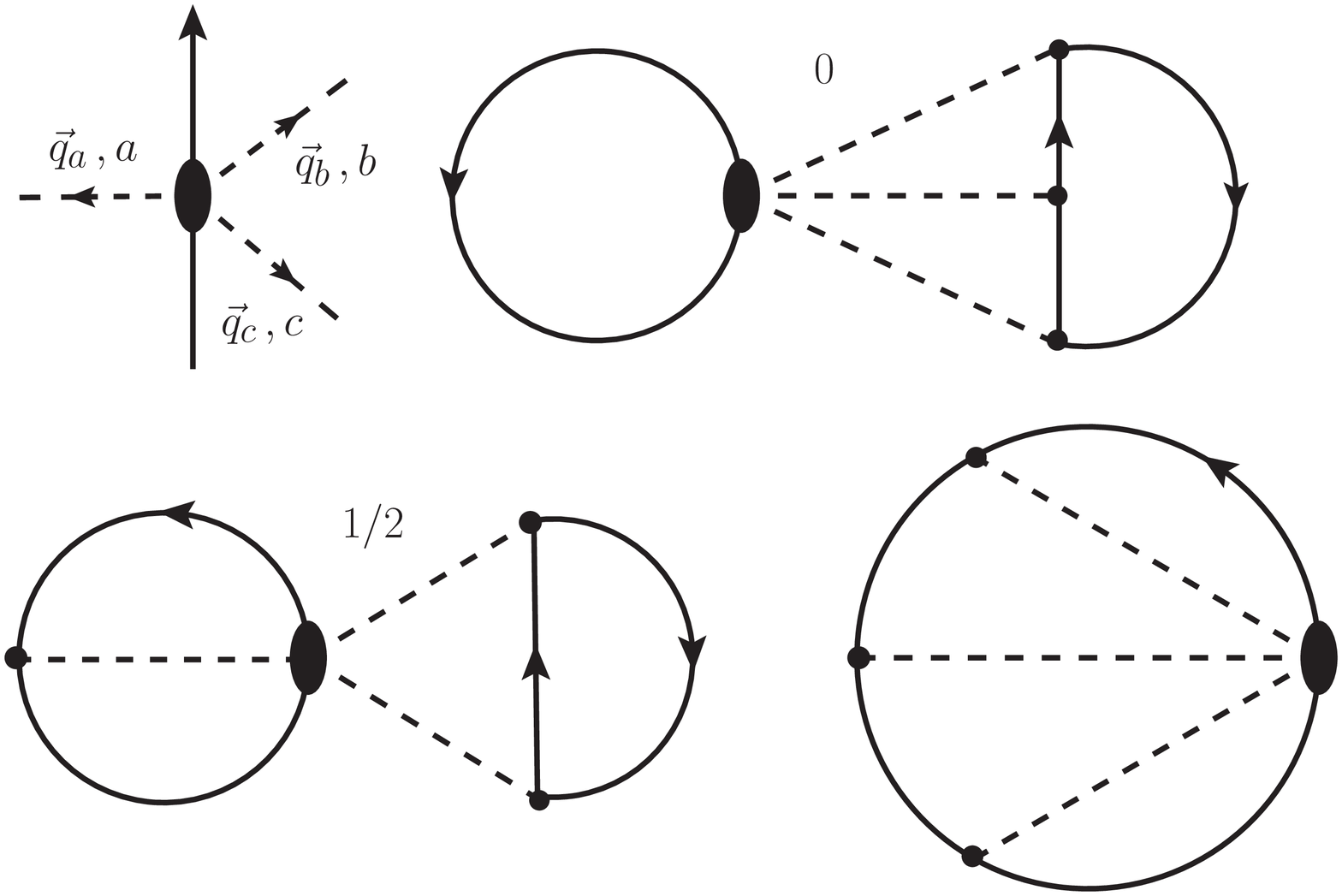}
\end{center}
\vspace{-.8cm}
\caption{The filled ellipse symbolizes the symmetrized three-pion 
contact-vertex involving double $\Delta$-excitation. The associated chiral 
four-nucleon interaction gives rise to one non-vanishing 2-ring diagram and  
one non-vanishing 1-ring diagram.}
\end{figure}

In this section we consider a further type of $\Delta$-induced long-range 
four-nucleon interaction \cite{deltuva}. Once the $\Delta$-isobar is excited, it 
can also couple directly to the pion (through the $\Delta\Delta\pi$-vertex). Each 
of the three pions emitted from the central nucleon gets absorbed on a spectator 
nucleon. For an efficient treatment of this chiral four-nucleon interaction it is 
advantageous to condense the three-pion dynamics at the central nucleon into 
a symmetrized three-pion contact-vertex as symbolized in Fig.\,10. Approximating 
the $\Delta$-propagator by the inverse mass-splitting $-i/ \Delta$ and summing 
over all six permutations of $(a,b,c)$ the pertinent NN$3\pi$ contact-vertex reads: 
\begin{eqnarray}&& {g_A^3 \over 40 f_\pi^3 \Delta^2} \bigg\{ -75 \epsilon_{abc} \,
\vec q_a\!\cdot\!(\vec q_b\!\times\!\vec q_c) + \vec q_a\!\cdot\! \vec q_b\,  
\vec \sigma\!\cdot \!\vec q_c\, (18 \delta_{ab}\tau_c-7  \delta_{ac}\tau_b-7  
\delta_{bc}\tau_a)\nonumber \\ && + \vec q_a\!\cdot\! \vec q_c\,  \vec 
\sigma\!\cdot\! \vec q_b\, (18 \delta_{ac}\tau_b-7  \delta_{ab}\tau_c-7\delta_{bc}
\tau_a)+ \vec q_b\!\cdot\!\vec q_c\,  \vec \sigma\!\cdot\! \vec q_a\, (18 
\delta_{bc}\tau_a-7 \delta_{ac}\tau_b-7\delta_{ab}\tau_c)\bigg\}\,, \end{eqnarray}
where $\vec q_a, \vec q_b$ and $\vec q_c$ denote outgoing pion-momenta. In order to 
fix the $\Delta\Delta\pi$-vertex we use the coupling constant ratio 
$g_{\pi \Delta \Delta}/g_{\pi NN}=1/5$ of the quark-model. The relation for isospin 
(transition) operators, $T_a \Theta_b T_c^\dagger = (5i\, \epsilon_{abc}-\delta_{ab}
\tau_c +4 \delta_{ac}\tau_b -\delta_{bc}\tau_a)/3$, with $\Theta_b$ the $4\times 4$ 
$\Delta$-isospin matrices has been employed together with an analogous relation for  
spin (transition) operators.

The 2-ring and 1-ring diagrams resulting from the three-pion contact-vertex are 
shown in Fig.\,10. The upper 2-ring diagram vanishes because the spin-trace (on the 
left) selects the term proportional to the scalar triple product from eq.(24) 
with linearly dependent vectors ($\vec q_a+\vec q_b+\vec q_c=\vec 0)$. For the lower 
2-ring diagram (with combinatoric factor $1/2$) the integrals over the Fermi sphere 
factorize and can be solved by using the master formula eq.(29). The corresponding 
contributions to the energy per particle of nuclear and neutron matter read:  
\begin{eqnarray}\bar E(\rho)&=& {5g_A^6 m_\pi^9 u\over 3(4\pi f_\pi)^6 \Delta^2}
\bigg[{4u^4\over 3}-6u^2+2+10u \arctan 2u-\bigg({9\over 2}+{1\over 2u^2}\bigg) 
\ln(1+4u^2)\bigg]\nonumber \\ &&\times  \bigg[12u^2-2-{16u^4\over 3}-16u 
\arctan 2u+\bigg(6+{1\over 2u^2}\bigg) \ln(1+4u^2)\bigg]\,, \end{eqnarray}
\begin{eqnarray}\bar E_n(\rho_n)&=& {g_A^6 m_\pi^9 u\over 18(4\pi f_\pi)^6 
\Delta^2}\bigg[{4u^4\over 3}-6u^2+2+10u \arctan 2u-\bigg({9\over 2}+{1\over 
2u^2}\bigg)\ln(1+4u^2)\bigg]\nonumber \\ &&\times  \bigg[12u^2-2-{16u^4\over 3}
-16u \arctan 2u+\bigg(6+{1\over 2u^2}\bigg) \ln(1+4u^2)\bigg]\,. \end{eqnarray}
The last 1-ring diagram in Fig.\,10 can be treated in the same way as 
the zig-zag diagram in Fig.\,8.  One finds the following double-integral 
representation for its contribution to the energy per particle of nuclear and 
neutron matter:
\begin{eqnarray}\bar E(\rho)&=& {3g_A^6 m_\pi^9 \over (4\pi f_\pi)^6
\Delta^2 u^3}\int_0^u\!dx\!\int_0^u\!dy \,\Bigg\{ {5\over 4} G_S(x) G_S(y)\Big[
4x y-L\Big] + {5\over 8}G_S(x) G_T(y) \nonumber \\ && \times \bigg[{3x \over y}
(1+x^2)- 5x y+{L\over 4}\bigg(2+6x^2 -3y^2-{3\over y^2}(1+x^2)^2\bigg) \bigg]  
\nonumber \\ && +{3\over 8}G_T(x) G_T(y) \bigg[ x y -{3x \over y}(1+x^2)+{L\over 
4} \bigg({3 \over y^2}(1+x^2)^2+2-3x^2\bigg)\bigg] \Bigg\}\,, \end{eqnarray}
\begin{eqnarray}\bar E_n(\rho_n)&=& {g_A^6 m_\pi^9 \over 24(4\pi f_\pi)^6\Delta^2 
u^3} \int_0^u\!dx\!\int_0^u\!dy \,\Bigg\{ G_S(x) G_S(y)\Big[4x y-L\Big] + 
G_S(x) G_T(y) \nonumber \\ && \times \bigg[10x y-{6x \over y}(1+x^2)
+L \bigg({3\over 2y^2}(1+x^2)^2-1-3x^2 +{3y^2\over 2}\bigg) \bigg]  \nonumber 
\\ &&+{1\over 5}G_T(x) G_T(y) \bigg[ 2x y -{6x \over y}(1+x^2)+L \bigg(1-
{3x^2\over 2}+{3 \over 2y^2}(1+x^2)^2\bigg)\bigg] \Bigg\}\,, \end{eqnarray}
with the logarithmic function $L$ defined in eq.(19). Note that the peculiar 
numerical coefficients entering the contact-vertex eq.(24), in particular the 
quark-model ratio $1/5$, have all disappeared in the final expressions for 
$\bar E(\rho)$ and $\bar E_n(\rho_n)$ in eqs.(25-28). 
\begin{figure}
\begin{center}
\includegraphics[scale=0.5,clip]{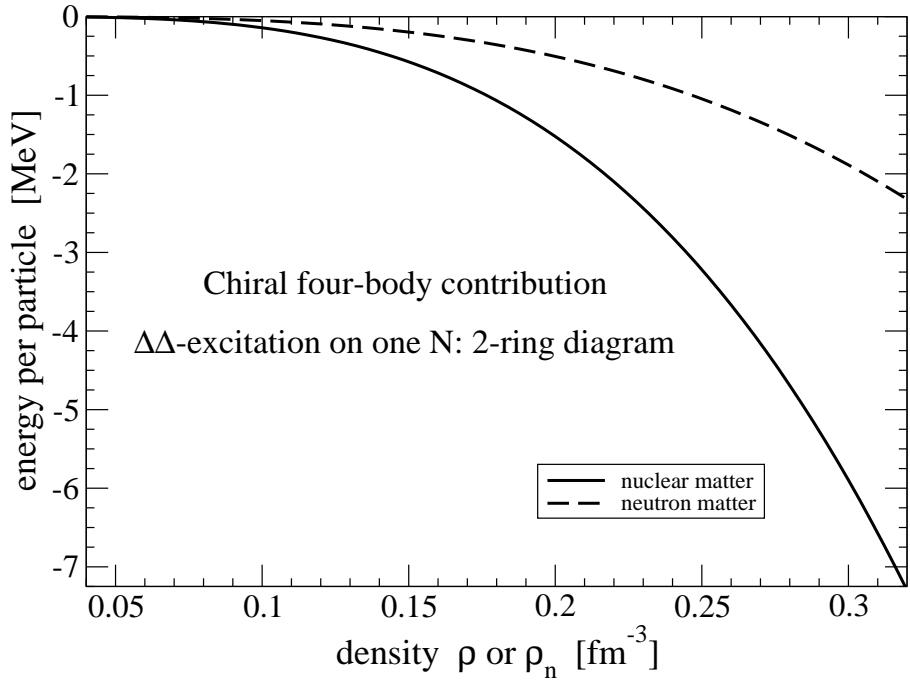}
\end{center}
\vspace{-.8cm}
\caption{$\Delta$-induced chiral four-body contributions to the energy per 
particle of nuclear and neutron matter: 2-ring diagram.}
\end{figure}

\begin{figure}
\begin{center}
\includegraphics[scale=0.5,clip]{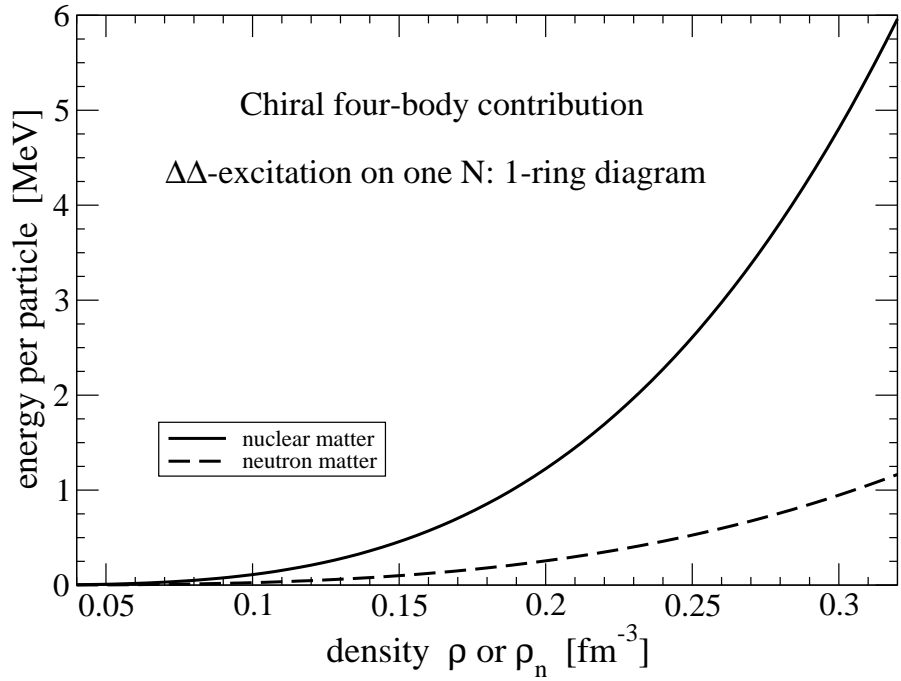}
\end{center}
\vspace{-.8cm}
\caption{$\Delta$-induced chiral four-body contributions to the energy per 
particle of nuclear and neutron matter: 1-ring diagram.}
\end{figure}

Numerical results for the energy per particle of nuclear and neutron matter
as they arise from the double $\Delta$-induced long-range four-nucleon interaction 
are shown in Figs.\,11,12. One observes attractive contributions from the 2-ring 
diagram which, in the next step, get nearly cancelled by repulsive contributions 
from the 1-ring diagram. After this balance the differences between nuclear and 
neutron matter which are individually visible in Figs.\,11,12 have essentially 
disappeared. The basic reason for the small net result in comparison to section 3 
is the absence of a 3-ring diagram (with its large spin-isospin weight factor) and 
the lower number of subsequent diagrams. The pattern of alternating and largely 
compensating 3-ring, 2-ring and 1-ring contributions shown in Figs.\,5,7,9,11,12 
is specific for the three-pion exchange four-nucleon interaction with two-fold 
$\Delta$-excitation considered here. When using fully antisymmetrized many-nucleon 
forces \cite{hebeler,achim} this merely technical decomposition into different 
ring diagrams does not come into play at all.         
\section{Combined results and discussion} 
\begin{figure}
\begin{center}
\includegraphics[scale=0.5,clip]{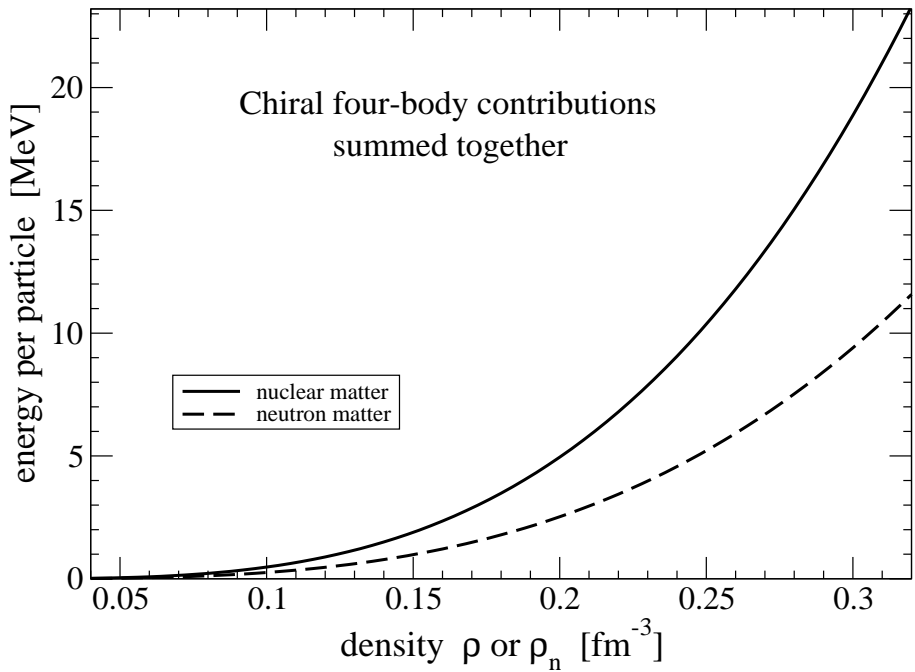}
\end{center}
\vspace{-.8cm}
\caption{Chiral four-body contributions to the energy per 
particle of nuclear and neutron matter summed together.}
\end{figure}

\begin{figure}
\begin{center}
\includegraphics[scale=0.5,clip]{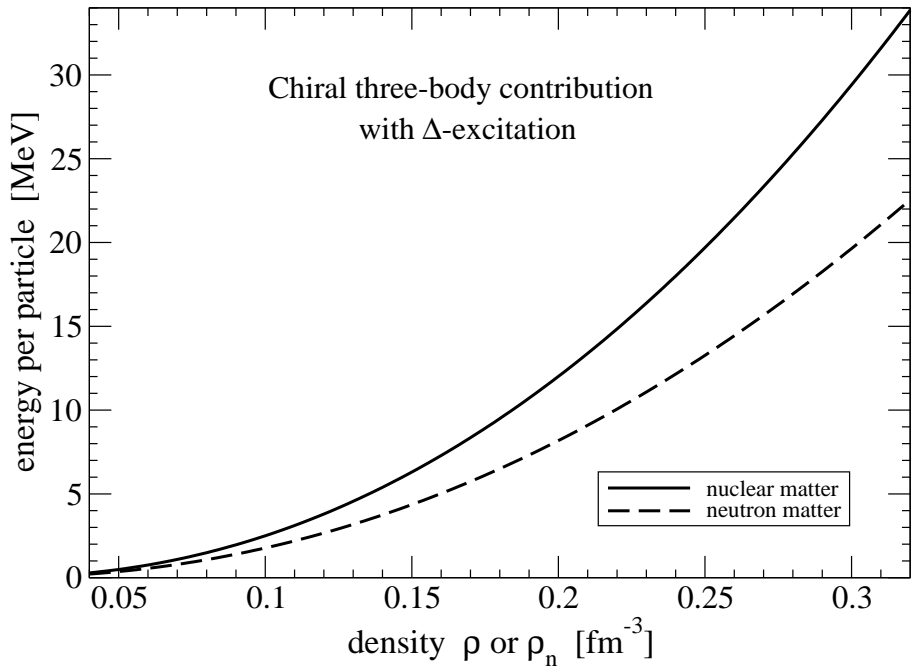}
\end{center}
\vspace{-.8cm}
\caption{Three-body contributions to the energy per particle of nuclear 
and neutron matter arising from two-pion-exchange with $\Delta$-isobar excitation.}
\end{figure}

After having evaluated various classes of long-range four-nucleon correlations  
in nuclear and neutron matter in the previous sections let us now draw a balance. 
The summed result of all these contributions is shown in Fig.\,13. For 
isospin-symmetric nuclear matter one finds at saturation density $\rho_0=
0.16\,$fm$^{-3}$ a moderate repulsive contribution to the energy per particle of 
$2.35\,$MeV. However, the corresponding curve for $\bar E(\rho)$ rises strongly 
with the density and reaches the value $23.3\,$MeV at $2\rho_0=0.32\,$fm$^{-3}$. 
One should note that in this density range, $0<\rho<2\rho_0$, a realistic nuclear 
matter equation of state resides in the binding regime with a negative energy 
per particle. A mere addition of long-range four-nucleon correlations 
as given by the full line in Fig.\,13 would produce enormous distortions.
In pure neutron matter the long-range four-body correlations come out about 
half as large (see dashed line in Fig.\,13). To put it into proportion, a value of 
$\bar E_n(2\rho_0)=11.6\,$MeV amounts to about $1/4$ of what sophisticated 
neutron matter calculations \cite{akmal} give at that density. It is also 
interesting to compare with the three-body correlations arising from $2\pi$-exchange 
with (single) $\Delta$-excitation in the same framework. The corresponding 
analytical formulas for $\bar E(\rho)$ and $\bar E_n(\rho_n)$ are written in 
eqs.(5,6,43,44) of ref.\cite{fritsch} and the associated numerical results are 
reproduced here in Fig.\,14. By comparison with Fig.\,13 one observes that for 
isospin-symmetric nuclear matter these long-range three-body correlations are 
at $2\rho_0$ about a factor $1.5$ larger than the four-body correlations, whereas in 
pure neutron matter the relative factor between both is about $2$. Note also that 
three-body and four-body correlations grow approximately quadratically and cubically 
with the density, respectively. The hierarchical order of multi-nucleon forces is 
therefore manifest for the long-range components considered in this work. Obviously, 
it gets more pronounced the lower the density ($\rho$ or $\rho_n$) is.

In a recent work by the Darmstadt-Ohio group \cite{achim} a calculation of the 
neutron matter equation of state $\bar E_n(\rho_n)$ with inclusion of the subleading 
chiral three-nucleon forces has been performed for the first time. These authors find 
relatively large attractive contributions, which reach about $-4\,$MeV at the highest 
considered density $\rho_n = 0.2\,$fm$^{-3}$ (see Fig.\,2 in ref.\cite{achim}). This 
value is to be compared with $2.5\,$MeV repulsion due to chiral four-body 
correlations studied in the present work. It is therefore conceivable that 
substantial cancellations will occur among these novel higher order correlations. It 
remains as a future task to demonstrate the expected cancellations at a quantitative 
level for pure neutron matter and also for isospin-symmetric nuclear matter.
    
\subsection*{Appendix: Reduction of integrals over Fermi spheres}
In this appendix we collect several formulas which are useful for reducing integrals 
over (multiple) Fermi spheres. If the integrand depends only on the magnitude of the 
momentum transfer, the following reduction formula holds:
\begin{equation} \int\limits_{|\vec p_{1,2}|<k_f}\!\!\!{d^3 p_1d^3 p_2\over (2\pi)^6} 
\, F(|\vec p_1\!-\!\vec p_2|) = {k_f^6 \over 3\pi^4} \int_0^1\! ds\, s^2(1-s)^2(2+s) 
F(2s k_f)\,.\end{equation} 
The weight function $s^2(1-s)^2(2+s)$ is slightly asymmetric about the midpoint 
$s=1/2$ and it reaches its maximum at $s = (\sqrt{4.2}-1)/2 = 0.5247$. These 
features are exhibited by the dashed line in Fig.\,15. In order to give a 
geometrical interpretation of this weight function, we note that the overlap 
volume of two unit-spheres with their centers displaced by $2s$ is $2\pi(1-s)^2(2+s)/3$.

Pions with their momentum-dependent interactions lead often to the following 
tensorial integral over a Fermi sphere of radius $k_f$:   
\begin{equation} \int\limits_{|\vec p_1|<k_f}\!\!\!{d^3 p_1\over (2\pi)^3} {(\vec p-
\vec p_1)_i (\vec p-\vec p_1)_j \over m_\pi^2+(\vec p-\vec p_1)^2}  = {m_\pi^3 
\over 48  \pi^2 x} \Big\{\delta_{ij} \, G_S(x,u) + (3\hat p_i
\hat p_j- \delta_{ij})\, G_T(x,u)  \Big\}\,.\end{equation} 
By contracting this equation with $\delta_{ij}$ and $3\hat p_i\hat p_j- \delta_{ij}$ 
the functions $G_S(x,u)$ and $G_T(x,u)$ can be calculated separately as simple scalar 
integrals over a Fermi sphere $|\vec p_1|<k_f$. The dimensionless variables on the 
right hand side are $x = |\vec p\,|/m_\pi$ and $u =k_f/m_\pi$. The explicit analytical 
expressions for the functions $G_{S,T}(x,u)$ involving arctangents and logarithms 
are given in eqs.(10,11), suppressing in the notation the second argument $u$.  

\begin{figure}
\begin{center}
\includegraphics[scale=0.5,clip]{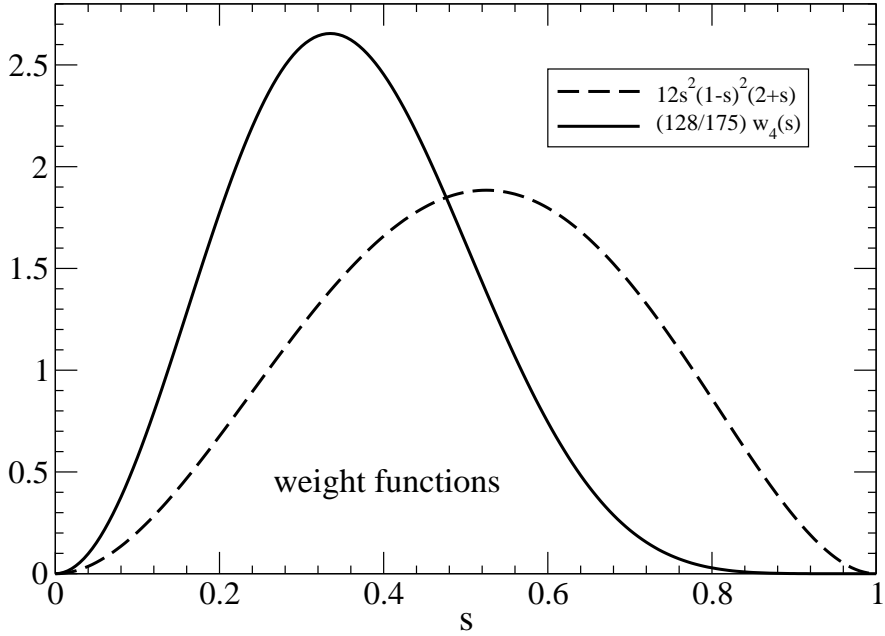}
\end{center}
\vspace{-.5cm}
\caption{Weight functions for integrals over two and four Fermi spheres, 
respectively. The area under both curves is equal to 1.}
\end{figure}

Let us add a further remarkable reduction formula for integrals over four Fermi 
spheres. We assume a dependence of the integrand on the magnitude of the total 
momentum $|\vec p_1\!+\!\vec p_2\!+\!\vec p_3\!+\!\vec p_4|$ only and derive the 
following reduction formula:   
\begin{equation} \int\limits_{|\vec p_j|<k_f}\!\!\!{d^{12} p\over (2\pi)^{12}} 
F(|\vec p_1\!+\!\vec p_2\!+\!\vec p_3\!+\!\vec p_4|) = {8 k_f^{12} \over 175 
(3\pi^2)^4} \int_0^1 \!ds\, w_4(s) F(4s k_f) \,,\end{equation} 
with the (non-smooth) weight function:
\begin{eqnarray} w_4(s) &=& 4s(1-s)^6(16 s^4+96s^3+156s^2+56s-9) \nonumber 
\\ && +\theta(1-2s) \, s(1-2s)^6(36+77s+24s^2-12s^3-4s^4) \,.\end{eqnarray} 
This peculiar result for $w_4(s)$ has been obtained with Fourier transformation 
techniques and eq.(31) has been checked numerically for many examples of $F(4s k_f)$.
By employing the Fourier representation of the delta-function $\delta^3(\vec p_1\!
+\!\vec p_2\!+\!\vec p_3\!+\!\vec p_4-4k_f \vec s\,)$ the weight function $w_4(s)$ 
results with a multiplicative factor $\pi/14175$ from the integral $s^2 \int_0^\infty 
dq \,q^2 j_0(4s q)[j_1(q)/q]^4$, where $j_{0,1}(q)$ denote spherical Bessel functions. 
The weight function $w_4(s)$ is normalized to  
$\int_0^1\!ds\, w_4(s) = 175/128$ and a quick analysis shows that it reaches its 
maximum value of $3.63$ at $s=0.335$. The asymptotic behavior is: $w_4(s) = 85 s^2$ 
for $s\to 0$, and $w_4(s) = 1260(1-s)^6$ for $s\to 1$. These feature are exhibited 
by the full line in Fig.\,15. The two weight functions put side by side in Fig.\,15 
demonstrate furthermore that multiple Fermi sphere integrals are dominated by 
total momenta of the order of the Fermi momentum $k_f$.

\subsection*{Acknowledgement}
I thank S. Fiorilla, A. Schwenk, W. Weise and A. Wirzba for valuable discussions 
and for specific advice.

\end{document}